\documentclass[12pt,oneside]{article}
\usepackage{amssymb,tabularx,multirow,amsmath,natbib,epsfig,threeparttable, amstext,subfigure,xcolor,bbm}
\usepackage{pdflscape}
\usepackage{afterpage}
\usepackage{capt-of}
\usepackage{rotating}
\usepackage[makeroom]{cancel}
\usepackage{bbm}
\usepackage{dsfont}

\usepackage{setspace} 
\usepackage[margin=1.0in]{geometry}

\usepackage[margin=20pt,labelfont={bf,small},textfont={it,small},indention=0.5cm]{caption}

\title{Multistage Hierarchical Capture-Recapture Models}

\author{Mevin B. Hooten$^{1}$, 
Michael R. Schwob$^{1}$, \\
Devin S. Johnson$^{3}$, and 
Jacob S. Ivan$^{4}$ \\
$^{1}$Department of Statistics and Data Sciences, \\ The University of Texas at Austin, \\ Austin, Texas, U.S.A. \\
$^{3}$Pacific Islands Fisheries Science Center, \\ National Marine Fisheries Service, \\National Oceanic and Atmospheric Administration, U.S.A.\\
$^{4}$Colorado Parks and Wildlife, Fort Collins, Colorado, U.S.A.}

\date{} 

\begin{document}

\maketitle

\begin{abstract}
Ecologists increasingly rely on Bayesian methods to fit capture-recapture models.  Capture-recapture models are used to estimate abundance while accounting for imperfect detectability in individual-level data.  A variety of implementations exist for such models, including integrated likelihood, parameter-expanded data augmentation, and combinations of those.  Capture-recapture models with latent random effects can be computationally intensive to fit using conventional Bayesian algorithms.  We identify alternative specifications of capture-recapture models by considering a conditional representation of the model structure.  The resulting alternative model can be specified in a way that leads to more stable computation and allows us to fit the desired model in stages while leveraging parallel computing resources.  Our model specification includes a component for the capture history of detected individuals and another component for the sample size which is random before observed.  We demonstrate this approach using three examples including simulation and two data sets resulting from capture-recapture studies of different species.        
\end{abstract}

\noindent%


\section{Introduction}
Formulations of statistical models for data arising from capture-recapture (CR) studies have evolved as our understanding of the models has improved, as technology has advanced, and as researchers have sought to extend them using hierarchical specifications.  In particular, Bayesian implementations of CR models remain popular for learning about wildlife population demographics \citep{royle2013spatial}.  Parameter-expanded data augmentation (PX-DA) approaches in Bayesian CR implementations have led to a variety of useful generalizations to accommodate heterogeneity.  The use of PX-DA is intuitive, but can increase computing requirements and could obscure other helpful model formulations.   

Fully hierarchical CR models implemented using standard Markov chain Monte Carlo (MCMC) methods \citep{royle2014hierarchical} often result in poorly mixed MCMC samples.  As a result, approaches based on integrated likelihoods are often sought \citep[e.g.,][]{efford2011estimation, yackulic2020need}.  Numerical integration approaches commonly used to fit CR models can be more stable \citep{bonner2014mc, king2016capture} but can also be computationally intensive.  We explore conditionally specified CR model formulations and consider recursive implementation strategies for fitting them.  These methods allow us to leverage parallel computing environments to fit CR models to data.  Also, conditional specification allows us to generalize CR models in ways that would not be apparent otherwise.  In what follows, we present the hierarchical capture-recapture model using parameter-expanded data augmentation and then show how a conditional specification of the model can be useful in recursive computing.  We show that alternative CR model specifications become apparent when fitting the model in stages: first while conditioning on the observed data ($\mathbf{y}_{1:n}$) and sample size ($n$) and second when updating the inference using the sample size itself         

Conventional CR models utilize data comprising binary detections of a subset of individual animals from a wildlife population when they are individually recognizable (either naturally or by artificial marking).  Bayesian CR models are often implemented using a parameter-expanded data augmentation (PX-DA) approach \citep{royle2007analysis, royle2012parameter}.  In this setting, individuals are observed over a set of sampling periods (or ``occasions'') $j=1,\ldots,J$, and binary detection/nondetection measurements $y_{i,j}$ for a set of $n$ observed individuals are recorded.  It is often assumed that $y_{i,j}$ are conditionally independent across occasions and the population is ``closed'' with respect to changes in demography and movement.  Thus, the count $y_{i}=\sum_{j=1}^J y_{i,j}$ represents the number of detections of individual $i$ with conditional mixture binomial distribution 
\begin{equation}
  y_{i} \sim 
  \begin{cases}
    \text{Binom}(J,p) &\mbox{, } z_i = 1 \\   
    \mathbbm{1}_{\{y_i=0\}} &\mbox{, } z_i = 0 \\   
  \end{cases} \;, 
  \label{eq:cr_datamodel}
\end{equation}
\noindent where $z_i$ is a latent population membership indicator for $i=1,\ldots,M$, with $M$ chosen so that it provides a realistic upper bound for population abundance (usually $M>>n$, where $n$ is the number of observed individuals).  In this type of PX-DA scenario, the data are augmented with all-zero capture histories such that $y_{i}=0$ for all $i=n+1,\ldots,M$ and the latent indicators are modeled as $z_i\sim \text{Bern}(\psi)$ \citep{royle2009analysis}.  After the data are observed, $z_i=1$ for $i=1,\ldots,n$ and the remaining $z_i$ for $i=n+1,\ldots,M$ are treated as unknown latent variables.   

In the Bayesian setting, priors are specified for the detection probability $p$ and membership probability $\psi$ in the homogeneous CR model in (\ref{eq:cr_datamodel}).  The PX-DA procedure induces a binomial process model on the total abundance of animals $N=\sum_{i=1}^M z_i$ such that $N\sim \text{Binom}(M,\psi)$.  When the prior for $\psi$ is uniform (i.e., $\psi\sim \text{Beta}(\alpha,\beta)$ with $\alpha=\beta=1$), it implies a discrete uniform prior for $N$ with support $\{0,1,\ldots,M\}$ when marginalized over $\psi$ (but see \citet{link2013cautionary} and \citet{villa2014cautionary} for varying perspectives on this choice of prior).   

Abundance models based on CR data have been generalized in a variety of ways, most of which are based on heterogeneity in detection probability such that $p_i$ is allowed to vary.  In some cases, $p_i$ is expressed as a function of environmental features or endogenous characteristics associated with individual $i$.  As CR models have been extended to accommodate spatially-explicit information that contributes to heterogeneity in $p_i$ such as distance between the center of an individual's activity region and the detector, they have also become more challenging to implement.

In what follows, we reformulate Bayesian CR models based on a conditional partitioning of the likelihood that is motivated by a multistage computing procedure.  This allows us to use recursive Bayesian computing methods to fit CR models to data.  These methods can facilitate implementation by allowing us to perform many of the necessary calculations in parallel between computing stages.

We review multistage Bayesian computing and then show how to apply it to the homogeneous CR model which suggests a way to formulate a broad class of capture-recapture models such that they are amenable to multistage computing strategies.  We extend the approach to heterogeneous CR models and demonstrate it using data collected as part of a study of salamanders in Great Smoky Mountains National Park and a study of snowshoe hares in central Colorado, USA.    

\section{Multistage Computing for Capture-Recapture}
We present a recursive Bayesian computing procedure that involves two or more stages to fit the CR model to data.  A variety of multistage computing strategies may be used with this framework including sequential Monte Carlo (SMC; \citealt{chopin2013smc2}) and the meta-analytic two-stage MCMC procedure described by \citet{lunn2013fully} that was generalized by \citet{hooten2019making}, where it was referred to as ``prior-proposal recursive Bayesian'' (PPRB) computation.  Recursive approaches to fitting certain classes of Bayesian ecological models have been demonstrated \cite[e.g.,][]{hooten2016hierarchical,gerber2018accounting, mccaslin2021tarb,feuka2022individual,leach2022recursive}, but have been less commonly used in Bayesian population modeling using capture-recapture data.

Our approach to multistage Bayesian computing relies on the ability to partition the data into two or more components and then write the posterior distribution as a product of conditional posterior distributions for each data partition given all those that were assimilated before it.  In the context of PX-DA and the conventional CR model with homogeneous detection probability that we presented in the previous section, the full posterior distribution can be written as 
\begin{equation}
  [p,\psi,\mathbf{z}_{(n+1):M}|\mathbf{y}_{1:M}]\propto \left(\prod_{i=1}^M [y_i|p,z_i][z_i|\psi]\right)[p][\psi]\;, 
  \label{eq:fullpost}
\end{equation}
\noindent where $z_i=1$ for $i=1,\ldots,n$ and $\mathbf{z}_{(n+1):M}=(z_{n+1},\ldots,z_M)'$ are unknown binary membership variables.  In this specification, we use bracket notation to denote probability distributions \citep{gelfand1990sampling} and express the conditional data model as
\begin{equation}
  [y_i|p,z_i]=z_i[y_i|p]+(1-z_i)\mathbbm{1}_{\{y_i=0\}} \;,
\end{equation}
\noindent where $[y_i|p]$ is a binomial probability mass function with $J$ trials and probability $p$ and the indicator function $\mathbbm{1}_{\{y_i=0\}}$ equals one when its condition is met and zero otherwise.  The product of data and process models can be marginalized over $z_i$ to yield the full likelihood component for individual $i$
\begin{equation}
  [y_i|p,\psi] = \psi[y_i|p]+(1-\psi)\mathbbm{1}_{\{y_i=0\}} \;, 
  \label{eq:intlik}
\end{equation}
\noindent which implies a mixture that is equivalent to a zero-inflated binomial model.  

Following \cite{king2016capture}, we partition the data into those that were observed $\mathbf{y}_{1:n}$ and those that were augmented $\mathbf{y}_{(n+1):M}$ (as zeros).  This allows us to express the posterior distribution from (\ref{eq:fullpost}) as proportional to (with respect to $p$ and $\psi$) the product of two terms  
\begin{equation}
  [p,\psi|\mathbf{y}_{1:n},\mathbf{y}_{(n+1):M},n]\propto [\mathbf{y}_{(n+1):M}|p,\psi,\mathbf{y}_{1:n},n][p,\psi|\mathbf{y}_{1:n},n]\;, 
  \label{eq:condpost}
\end{equation}
\noindent where we have used the marginalized data model in (\ref{eq:intlik}) to reduce notation with respect to the latent variables $z_i$.  Critically, our partitioning scheme depends on $n$ (the number of observed individuals).  Conditioning on $n$ implies that we know which of the $M$ possible individuals in the superpopulation were observed.  Thus, the second term on the right-hand side of (\ref{eq:condpost}) can be written as
\begin{equation}
  [p,\psi|\mathbf{y}_{1:n},n]\propto \left(\prod_{i=1}^n [y_i|p,y_i>0]\right)[n|p,\psi][p][\psi] \;,
  \label{eq:M0firstpost}
\end{equation}
\noindent where the conditional data model in (\ref{eq:M0firstpost}) is a zero-truncated binomial with PMF
\begin{equation}
  [y_i |p,y_i>0] \propto \frac{[y_i|p]}{1-(1-p)^J} \;,
  \label{eq:ztbinom}
\end{equation}
\noindent such that $[y_i|p]$ is a binomial probability mass function.   

The number of observed individuals is the sum of individual-level indicator variables that express which individuals were detected/undetected as $n=\sum_{i=1}^M\mathbbm{1}_{\{y_i>0\}}$.  Conditional on $p$ and $\psi$, the probability of detecting an individual from the superpopulation is  
\begin{equation}
  \text{Pr}(\mathbbm{1}_{\{y_i>0\}}=1 |p,\psi) = \psi (1-(1-p)^J) \;,  
\end{equation}
\noindent and $\mathbbm{1}_{\{y_i>0\}}$ are conditionally independent Bernoulli trials before $\mathbf{y}_{1:n}$ are observed.  Thus, the number of observed individuals is distributed conditionally as $n\sim [n|p,\psi]=\text{Binom}(M,\psi(1-(1-p)^J)$ under this model.    

In most other recursive implementations, we would need to evaluate the conditional data distribution $[\mathbf{y}_{(n+1):M}|p,\psi,\mathbf{y}_{1:n},n]$ in the first term on the right-hand side of (\ref{eq:condpost}).  However, in this PX-DA situation where $\mathbf{y}_{(n+1):M}=\mathbf{0}$ are augmented data, the conditional data distribution is proportional to $\prod_{i=n+1}^M \mathbbm{1}_{\{y_i=0\}}=1$.  Thus, we do not need to consider the augmented data $\mathbf{y}_{(n+1):M}$ in this framework, only the observed data $\mathbf{y}_{1:n}$ and the sample size $n$.  The data augmentation merely suggests that we can reformulate the homogeneous CR model as in (\ref{eq:M0firstpost}). 

 Specifications like that shown in (\ref{eq:M0firstpost}) have associated multistage computing strategies.  In our case, the specification involves a model for the observed data conditioned on $n$ and the model parameters (e.g., $p$ and $\psi$) and also a model for $n$ conditioned on the parameters.  Similar CR model specifications were discussed by \cite{borchers2008spatially} and differ from capture-recapture model specifications that are based on $N$, the total abundance, directly \citep[e.g.,][]{king2016capture}.  Neither \cite{borchers2008spatially} nor \cite{king2016capture} leveraged the formulation to facilitate recursive Bayesian computing. 

In the case of the homogeneous CR model we described, we can fit it recursively in the following way.  For the first stage, we fit a model using Bayesian methods and a stochastic sampling procedure (e.g., importance sampling, MCMC, Hamiltonian Monte Carlo) based on the conditional data model and priors for $p$ and $\psi$ in (\ref{eq:M0firstpost}) comprising a temporary posterior distribution proportional to 
\begin{equation}
  \left(\prod_{i=1}^n [y_i|p,y_i>0]\right)[p][\psi] \;.
  \label{eq:temp_post}
\end{equation}

Then, in the second stage of the procedure, we use a randomly selected first-stage sample $(p^{(*)},\psi^{(*)})$ as a proposal in either an importance \citep[e.g.,][]{chopin2002sequential} or Metropolis-Hastings ratio depending on whether SMC or MCMC is preferred for the second stage.  For MCMC, the second-stage Metropolis-Hastings ratio can be written as    
\begin{equation}
  r=\frac{[n|p^{(*)},\psi^{(*)}]}{[n|p^{(k-1)},\psi^{(k-1)}]} \;, 
  \label{eq:mhratio}
\end{equation}
\noindent for MCMC iteration $k$ and we let $p^{(k)}=p^{(*)}$ and $\psi^{(k)}=\psi^{(*)}$ with probability $\text{min}(r,1)$; we retain $p^{(k)}=p^{(k-1)}$ and $\psi^{(k)}=\psi^{(k-1)}$ otherwise (Supporting Information, Appendix A; \citealt{hooten2019making}).  The second stage does not involve tuning parameters and is unsupervised.  

After the first stage, the resulting sample for the model parameters $p$ and $\psi$ can be used to compute all possible numerators (and hence also denominators) of (\ref{eq:mhratio}) in parallel.  For the homogeneous CR model, a parallelized intermediate computing step may not be necessary because the conditional mass function of $n$ can be evaluated quickly.  However, in the heterogeneous CR models that follow, we need to use numerical or stochastic integration techniques to compute the components of the Metropolis-Hastings ratio (\ref{eq:mhratio}) and parallelization leads to substantial reductions in computing time.   

In the PX-DA framework based on the hierarchical CR model, $N=\sum_{i=1}^M z_i$ is treated as a derived quantity.  In a recursive framework, we obtain an MCMC sample for $N$ in a third computing stage where $N_0$ is sampled from its full-conditional distribution as $N_0^{(k)}\sim \text{Binom}(M-n,\psi^{(k)}(1-p^{(k)})^J/(\psi^{(k)}(1-p^{(k)})^J+1-\psi^{(k)}))$ and then $N^{(k)}=n+N_0^{(k)}$ for MCMC iteration $k=1,\ldots,K$.  The quantity $N_0$ represents the undetected individuals from the superpopulation that were part of our study population.  Critically, our model fit does not depend on this third stage, but the inference we obtain about abundance $N$ as a derived quantity is fully Bayesian.  This is in contrast to the ``empirical Bayes'' estimator proposed by \citet{dorazio2013bayes} which depends on a point estimate of $N_0$ that is based on maximum likelihood estimators of the parameters. 

We demonstrate the application of this approach to multistage Bayesian computing by fitting the hierarchical CR model to simulated data in the Supporting Information, Appendix B.  We compared inference from both the conventional single-stage MCMC algorithm and the PPRB procedure and illustrated their equivalence.   

\section{Generalizations and Alternative Models for $n$}
An important aspect of expressing the posterior distribution as in (\ref{eq:M0firstpost}) is that it admits other specifications of the model for the number of observed individuals $n$.  In the previous section, we derived a conditional binomial distribution for $n$ that is consistent with the assumptions of the original PX-DA implementation of the homogeneous CR model.  Alternatively, our reformulation of the model allows us to specify any conditional model for $n$; for example, we could specify a conditional Poisson model for $n$ instead.  Similar Poisson models for $n$ (and $N$) have been suggested previously \citep[e.g.,][]{borchers2008spatially,johnson2010model,dorazio2013bayes,schofield2014hierarchical,king2016capture}.  It is well-known that as $M\rightarrow\infty$ (common in PX-DA implementations of CR models) and $\psi\rightarrow 0$ such that $\lambda=\psi M$ is constant, then the binomial converges in distribution to the Poisson such that $n\sim \text{Pois}(\lambda (1-(1-p)^J))$.  Critically, the true generating mechanisms are not known for real data and thus the flexibility we gain by generalizing the conditional model for $n$ can help accommodate a wider range of scenarios.  Furthermore, certain specifications  of the conditional model for $n$ may facilitate the implementation of CR models, as we show in the examples that follow.  

Consider the hierarchical CR model with heterogeneous detection probability that varies by individual and is implemented using PX-DA such that $\mathbf{y}_{1:n}$ are observed counts of detections for a set of $n$ individuals and $\mathbf{y}_{(n+1):M}=\mathbf{0}$ represent the augmented individuals \citep{royle2008hierarchical,schofield2014hierarchical}.  The conventional hierarchical specification for this heterogeneous CR model is $y_i\sim \psi [y_i|p_i] + (1-\psi)\mathbbm{1}_{\{y_i=0\}}$ with the individual-specific detection probabilities modeled as $\text{logit}(p_i)\sim \text{N}(\mu,\sigma^2)$ for $i=1,\ldots,M$.  This model treats the detection probabilities as random effects and thus requires priors $[\mu]$ and $[\sigma^2]$.  

The heterogeneous CR model is a good candidate for recursive computing strategies because it can be challenging to implement \citep{king2016capture, white2017population}.  It can be reformulated as described in the previous section based on a recursive implementation where we condition on $n$ and $p_i$ which yields the zero-truncated binomial data model for positive counts $y_i\sim [y_i|p_i]/(1-(1-p_i)^J)$ for $i=1,\ldots,n$ \citep{borchers2008spatially}.  However, to improve stability of the first-stage algorithm, we marginalize over $\mathbf{p}$ and write the associated posterior distribution for this model as
\begin{equation}
  [\mu,\sigma^2,\psi|\mathbf{y}_{1:n},n]\propto [\mathbf{y}_{1:n}|\mu,\sigma^2,n][n|\mu,\sigma^2,\psi][\mu][\sigma^2][\psi] \;,
  \label{eq:Mhpost}
\end{equation}
\noindent where we describe its components in what follows.  The integrated data model in (\ref{eq:Mhpost}) can be expressed as 
\begin{equation}
  [\mathbf{y}_{1:n}|\mu,\sigma^2,n]=\prod_{i=1}^n \int [y_i|p_i,y_i>0][\text{logit}(p_i)|\mu,\sigma^2,y_i>0]d\text{logit}(p_i), 
  \label{eq:Mhconddatamod}
\end{equation}
\noindent which is the probability of observing capture histories $\mathbf{y}_{1:n}$ given those $n$ individuals were observed.  We described the conditional data model $[y_i|p_i,y_i>0]$ in (\ref{eq:ztbinom}) and because of the heterogeneity, we also need to condition on $y_i>0$ in the distribution for $\text{logit}(p_i)$. Thus, we write the conditional process model in (\ref{eq:Mhconddatamod}) 
\begin{align}
  [\text{logit}(p_i)|\mu,\sigma^2,y_i>0]&=\frac{\text{Pr}(y_i>0|\text{logit}(p_i),\mu,\sigma^2)[\text{logit}(p_i)|\mu,\sigma^2]}{\int \text{Pr}(y_i>0|\text{logit}(p_i),\mu,\sigma^2)[\text{logit}(p_i)|\mu,\sigma^2] d\text{logit}(p_i)} \;, \\ 
  & =\frac{(1-(1-p_i)^J)[\text{logit}(p_i)|\mu,\sigma^2]}{\int (1-(1-p)^J)[\text{logit}(p)|\mu,\sigma^2] d\text{logit}(p)} \;,
\end{align}
\noindent which allows us to rewrite (\ref{eq:Mhconddatamod}) as 
\begin{equation}
  [\mathbf{y}_{1:n}|\mu,\sigma^2,n]=\frac{\prod_{i=1}^n \int [y_i|p_i][\text{logit}(p_i)|\mu,\sigma^2]d\text{logit}(p_i)}{\left(\int (1-(1-p)^J)[\text{logit}(p)|\mu,\sigma^2] d\text{logit}(p)\right)^n} \;. 
  \label{eq:Mhconddatamod2}
\end{equation}

The integrated conditional distribution for $n$ from the joint distribution in (\ref{eq:Mhpost}) can be calculated as the $M$-dimensional integral 
\begin{equation}
   [n|\mu,\sigma^2,\psi]=\int [n|\mathbf{p}_{1:M},\psi][\text{logit}(\mathbf{p}_{1:M})|\mu,\sigma^2]d\text{logit}(\mathbf{p}_{1:M}) \;,
   \label{eq:npartialint}
\end{equation}
\noindent where $[\text{logit}(\mathbf{p}_{1:M})|\mu,\sigma^2]=\prod_{i=1}^M[\text{logit}(p_i)|\mu,\sigma^2]$ and the conditional distribution $[n|\mathbf{p}_{1:M},\psi]$ is Poisson-binomial with $M$ trials and probabilities $\psi (1-(1-p_i)^J)$ for $i=1,\ldots,M$ \citep{fernandez2010closed}, as implied by the conventional heterogeneous CR model.  For reference, a Poisson-binomial distribution represents the sum of independent Bernoulli random variables, each with its own success probability. However, the Poisson-binomial PMF is numerically inefficient to calculate, often requiring a Fourier transform approach.  

Alternatively, we could specify that each detection indicator is Poisson distributed with intensity $\psi(1-(1-p_i)^J)$.  Then, if we assume large $M$ and conditionally indepenent detections, the sum $n=\sum_{i=1}^M \mathbbm{1}_{\{y_i>0\}}$ can be modeled as $n\sim \text{Pois}(\psi\sum_{i=1}^M (1-(1-p_i)^J))$.  The Poisson PMF is much more numerically tractable and leads to a faster numerical approximation of (\ref{eq:npartialint}).  

To fit the model with the posterior distribution in (\ref{eq:Mhpost}) using the PPRB approach, we first obtain a sample from the posterior distribution associated with the observed data $\mathbf{y}_{1:n}$ and then assimilate the number of observed individuals $n$ in a second stage.  There are several important implementation details that arise for this heterogeneous CR model.  In a first computing stage, we use a standard algorithm to obtain a MCMC sample based on the joint distribution $[\mathbf{y}_{1:n}|\mu,\sigma^2,n][\mu][\sigma^2][\psi]$,
which requires Metropolis-Hastings updates for $\mu$ and $\sigma^2$.  We can obtain a Monte Carlo sample for $\psi$ from its prior because it does not appear in the conditional data model.

Using the MCMC sample resulting from the first stage, we evaluate $[n|\mu,\sigma^{2},\psi]$ for all realizations of the parameters in parallel.  This intermediate computing step approximates the integrated PMF for $n$ using Monte Carlo integration based on (\ref{eq:npartialint}), or other numerical approach (see \citealt{king2016capture} for a quadrature method). 

Then, in the second computing stage, we use a random draw from the first stage MCMC sample as the proposal $\{\mu^{(*)},\sigma^{2(*)},\psi^{(*)}\}$ and update using the Metropolis-Hastings ratio
\begin{equation}
  r=\frac{[n|\mu^{(*)},\sigma^{2(*)},\psi^{(*)}]}{[n|\mu^{(k-1)},\sigma^{2(k-1)},\psi^{(k-1)}]} \;, 
  \label{eq:Mh_mh}
\end{equation}
\noindent where we use the relevant numerator in (\ref{eq:Mh_mh}) from the parallel computing output using a look-up table.  This parallel computing procedure is similar to pre-fetching \citep{brockwell2006parallel}, but for the entire Markov chain.

In a third computing stage, we sample the population abundance parameter using the MCMC output from the second stage and an approach similar to what we described for the homogeneous model.  Thus, to obtain a MCMC sample for $N$ based on the heterogeneous model, we draw $N_0^{(k)}$ from its full-conditional distribution and let $N^{(k)}=n+N_0^{(k)}$. 
 
For example, if we specified our conditional model $[n|\mathbf{p}_{1:M},\psi]$ as $\text{Pois}(\psi\sum_{i=1}^M(1-(1-p_i)^J))$, then we obtain a MCMC sample for $N$ by drawing  
\begin{equation}
  N_0^{(k)} \sim \text{Pois}(\bar{\psi}^{(k)}(M-n)) \;,
\end{equation}
\noindent where the term $\bar{\psi}^{(k)}$ is the full-conditional probability of population membership for an augmented individual averaged over the conditional distribution of $\text{logit}(p)$ and is homogeneous for all $i=n+1,\ldots,M$.  We calculate $\bar{\psi}^{(k)}$ as
\begin{equation}
  \bar{\psi}^{(k)}=\int \left(\frac{\psi^{(k)}(1-p)^J}{\psi^{(k)}(1-p)^J+1-\psi^{(k)}}\right) [\text{logit}(p)|\mu^{(k)},\sigma^{2(k)}]d\text{logit}(p) \;,
\end{equation}
\noindent for $k=1,\ldots,K$ second-stage MCMC iterations.

To assess the sampling strategy and study design, it is common to infer the ``power to detect,'' which is the probability of detecting a randomly selected individual from the population in $J$ sampling occasions \citep[e.g.,][]{dupont2021optimal}.  In closed-population models with homogeneous detection probability $p$, the conditional power to detect is calculated as $\text{Pr}(\tilde{y}>0|\tilde{z}=1,p)=1-(1-p)^J$.  This quantity can be readily extended to Bayesian CR models with heterogeneous detectability by considering the posterior power to detect  
\begin{align}
  \text{Pr}(\tilde{y}>0|\tilde{z}=1,\mathbf{y})&=\int \mathbbm{1}_{\{\tilde{y}>0\}} [\tilde{y}|\tilde{z}=1,\mathbf{y}]d\tilde{y} \;, \\
  &=\int\int\int\int  \mathbbm{1}_{\{\tilde{y}>0\}} [\tilde{y}|\tilde{p}][\text{logit}(\tilde{p})|\mu,\sigma^2][\mu,\sigma^2|\mathbf{y}]d\text{logit}(\tilde{p})d\mu d\sigma^2 d\tilde{y} \;,
  \label{eq:ptd}
\end{align}
\noindent which is a derived posterior predictive quantity.  We can use composition sampling to obtain a MCMC sample $\tilde{y}^{(k)}$ for $k=1,\ldots,K$ and then Monte Carlo integration to approximate the posterior power to detect as $\text{Pr}(\tilde{y}>0|\tilde{z}=1,\mathbf{y})=\sum_{k=1}^K \mathbbm{1}_{\{\tilde{y}^{(k)}>0\}}/K$.  We can also compute $E(n/N|\mathbf{y})$ as an alternative way to represent power to detect.   

\subsection{Application:  Salamander abundance}
We demonstrate the PPRB approach to implementing the heterogeneous CR model using a data set comprised of encounter histories based on $J=4$ sampling occasions for red-cheeked salamander (\emph{Plethodon} \emph{jordani}).  These data were collected in Great Smoky Mountains National Park in a 15 m $\times$ 15 m fenced plot to ensure closure of the population under study \citep{bailey2004spatial,hooten2019bringing}.  The measurement process resulted in $n=93$ observed individuals with 78, 11, and 4 detected on 1, 2, and 3 sampling occasions, respectively.  This species is known to have low detectability, thus we augmented the data with $M-n=1407$ all-zero encounter histories, which implies $M=1500$ total individuals in our superpopulation.        

A variety of factors can result in heterogeneous capture probabilities for red-cheeked salamanders \citep{bailey2004spatial}.  To account for individually varying detectability, we fit heterogeneous CR models to these data using the two-stage PPRB procedure.  For comparison, we fit two heterogeneous CR models; one based on the conditional Poisson-binomial assumption for $n$ and the other based on the conditional Poisson assumption for $n$.  We specified priors for both models as:  $\mu \sim \text{N}(-1,1)$, $\sigma^2\sim \text{IG}(0.01,0.01)$, and $\psi\sim\text{Beta}(1,1)$.  

We fit the models using $K=500000$ MCMC iterations on a 28-core machine with 2.5 Ghz processors.  The first stage algorithm required approximately 16.2 minutes.  The second stage algorithms required 9.3 hours for the Poisson-binomial version and only 26.2 minutes for the Poisson version.  For comparison, the time per effective MCMC sample was $0.037$ minutes for the Poisson-binomial, $0.003$ for the Poisson, and $0.107$ for a single-stage implementation of the model (using JAGS; \citealt{plummer2003jags}).  These results imply the recursive Poisson implementation was two orders of magnitude faster per effective sample size than the single-stage algorithm.  This highlights an advantage in having the flexibility to generalize the model specification by modifying the conditional distribution for $n$. 

\begin{figure}[htp]
  \centering
  \includegraphics[width=4in]{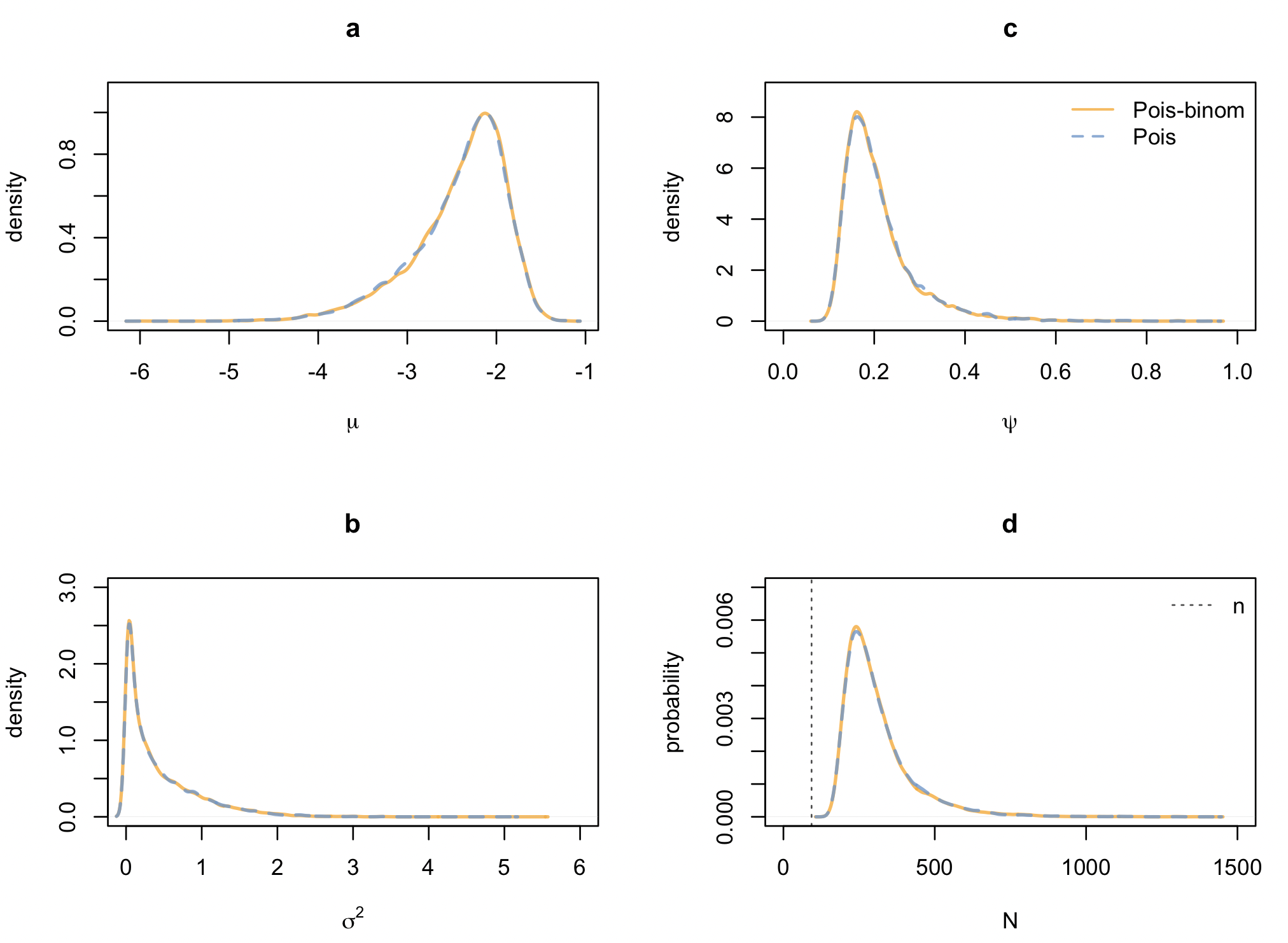}
  \caption{Marginal posterior distributions for a) $\mu$, b) $\sigma^2$, c) $\psi$, and d) $N$.  Distributions shown are a result of the multistage Bayesian algorithms associated with the model with Poisson-binomial (solid orange) and Poisson (dashed blue) assumption for $n$.  Subfigure d shows marginal posterior probability mass functions for $N$ as smoothed lines for comparison purposes due to the extent of the support. Priors shown as dashed gray lines.}
  \label{fig:Mh_results}
\end{figure}

The posterior results are summarized in Figure~\ref{fig:Mh_results}. Fitting the two heterogeneous CR models results in remarkably similar posterior distributions for the parameters $\mu$, $\sigma^2$, and $\psi$ (Figure~\ref{fig:Mh_results}a-c). 
The inference for abundance $N$ was also similar for the two models (Figure~\ref{fig:Mh_results}d).  The conventional heterogeneous CR model with Poisson-binomial distribution for $n$ had a slightly smaller posterior mean abundance ($E(N|\mathbf{y})=310.8$) than the model with Poisson distribution for $n$ ($E(N|\mathbf{y})=312.3$) but identical 95\% credible interval $[181, 626]$.  Similarly, for both models, the posterior power to detect was approximately $\text{Pr}(\tilde{y}>0|\tilde{z}=1,\mathbf{y})=0.33$ for this species, based on $J=4$ sampling occasions.  This result was the same as $E(n/N|\mathbf{y})$ numerically and implies that we have a $33\%$ chance of detecting a randomly selected individual in the population. 

\section{Multistage Computing for Spatial Capture-Recapture}
We can apply the multistage computing procedure described in previous sections to fit spatial CR (SCR; \citealt{royleyoung2008hierarchical}) models.  In the spatially explicit setting, individuals may be detected at an array of ``traps'' (i.e., detectors) located at positions $\mathbf{x}_l$ for $l=1,\ldots,L$.  Thus, we retain the CR data model from before 
\begin{equation}
  y_{i,l} \sim 
  \begin{cases}
    \text{Binom}(J,p_{i,l}) &\mbox{, } z_i = 1 \\   
    \mathbbm{1}_{\{y_{i,l}=0\}} &\mbox{, } z_i = 0 \\   
  \end{cases} \;, 
  \label{eq:scr_datamodel}
\end{equation}
\noindent for $i=1,\ldots,n,n+1,\ldots,M$ and where $z_i\sim \text{Bern}(\psi)$ are binary variables indicating population membership as before.  Using PX-DA, the observed data are augmented with all-zero capture histories such that $y_{i,l}=0$ for $i=n+1,\ldots,M$ and $l=1,\ldots,L$.  In SCR, heterogeneity in detection/capture probability is often characterized as a function of the distance between (unknown) individual-based activity centers $\mathbf{s}_i$ and the (known) trap locations $\mathbf{x}_l$.  For example, we can use a logit link function such that      
\begin{equation}
  \text{logit}(p_{i,l})=\beta_0 + \beta_1 ||\mathbf{s}_i-\mathbf{x}_l||^2_2 \;,
\end{equation}
\noindent where we treat $\mathbf{s}_i$ as random effects with distribution $\mathbf{s}_i\sim [\mathbf{s}]$.  We note that alternative link functions (e.g., `cloglog') are also popular in SCR models \citep[e.g.,][]{royle2008hierarchical,hooten2019bringing}.  In many cases, the distribution $[\mathbf{s}]$ serves as a prior and is specified as a bivariate uniform distribution over the study area, which implies a complete spatial random point process for $\mathbf{s}_i$ before the data are observed.  Various approaches have been proposed to generalize the model for $\mathbf{s}_i$ such as allowing it to be a heterogeneous spatial point process \citep[e.g.,][]{sutherland2015modelling}.      
 
Following the procedure we described in the previous section, we express the posterior distribution associated with the SCR model as 
\begin{equation}
  [\boldsymbol\beta,\psi|\mathbf{Y}_{1:n},n]\propto \left(\prod_{i=1}^n \left[\mathbf{y}_i \Bigm\vert \boldsymbol\beta,\sum_{l=1}^L y_{i,l}>0\right]\right)[n|\boldsymbol\beta,\psi][\boldsymbol\beta][\psi] \;,
  \label{eq:SCRpost}
\end{equation}
\noindent and describe its components in what follows.  The integrated data model in (\ref{eq:SCRpost}) can be expressed as  
\begin{equation}
  \left[\mathbf{y}_i \Bigm\vert \boldsymbol\beta,\sum_{l=1}^L y_{i,l}>0\right]=\frac{\int [\mathbf{y}_i|\mathbf{p}_i][\mathbf{s}_i] d\mathbf{s}_i}{\int (1-\prod_{l=1}^L (1-p_{i,l})^J)[\mathbf{s}_i]d\mathbf{s}_i}
\end{equation}
\noindent using the same arguments as in the previous section, where $[\mathbf{y}_i|\mathbf{p}_i]=\prod_{l=1}^L [y_{i,l}|p_{i,l}]$ is a product over binomial PMFs with $J$ trials. 

The conditional distribution of $n$ is 
\begin{equation}
   [n|\boldsymbol\beta,\psi]=\int [n|\mathbf{S}_{1:M},\psi][\mathbf{S}_{1:M}]d\mathbf{S}_{1:M} \;,
   \label{eq:npartialintSCR}
\end{equation}
where the joint prior for the activity centers is $[\mathbf{S}_{1:M}]=\prod_{i=1}^M[\mathbf{s}_i]$ and $[n|\mathbf{S}_{1:M},\psi]$ is either Poisson-binomial under the conventional PX-DA model specification or, as we demonstrate in the example that follows, Poisson with intensity $\sum_{i=1}^M \psi(1-\prod_{l=1}^L(1-p_{i,l})^J)$ such that $p_{i,l}=\text{logit}^{-1}(\beta_0+\beta_1||\mathbf{s}_i-\mathbf{x}_l||^2_2)$.  This Poisson intensity is derived as before by considering the number of observed individuals as a sum of detection indicators.  In the context of our SCR model, $n=\sum_{i=1}^M \mathbbm{1}_{\{\sum_{l=1}^L y_{i,l}>0\}}$, where each $\mathbbm{1}_{\{\sum_{l=1}^L y_{i,l}>0\}}$ is a binary random variable with success probability
\begin{align}
  P(\mathbbm{1}_{\{\sum_{l=1}^L y_{i,l}>0\}}=1|\mathbf{p}_i,\psi)&=1-P(\mathbbm{1}_{\{\sum_{l=1}^L y_{i,l}>0\}}=0|\mathbf{p}_i,\psi) \;, \\ 
  &=1-\left(\psi\prod_{l=1}^L P(y_{i,l}=0|p_{i,l}) +1-\psi\right)\;, \\
  &=\psi\left(1-\prod_{l=1}^L (1-p_{i,l})^J\right) \;. \label{eq:scr_prob} 
\end{align} 
\noindent We note that (\ref{eq:scr_prob}) reduces to the nonspatial heterogeneous CR probability of detection when there is a single trap ($L=1$).  Thus, assuming large $M$ and conditional independence, we treat each probability in (\ref{eq:scr_prob}) as an intensity and sum across $M$ individuals in the superpopulation to obtain the total Poisson intensity for $n$.  Alternatively, for the Poisson-binomial model with $M$ trials, we use (\ref{eq:scr_prob}) as probabilities for $i=1,\ldots,M$.     

To fit this version of the SCR model using recursive Bayesian computing strategies, we use the same sequence of stages described in the previous section.  In stage 1, we use MCMC to fit the model with joint distribution 
\begin{equation}
 \left(\prod_{i=1}^n \left[\mathbf{y}_i \Bigm\vert \boldsymbol\beta,\sum_{l=1}^L y_{i,l}>0\right]\right)[\boldsymbol\beta][\psi] \;, 
 \label{eq:SCRfirstpost}
\end{equation}
\noindent which involves Metropolis-Hastings updates for $\boldsymbol\beta$, but direct Monte Carlo sampling for $\psi$ from its prior because it does not appear in the integrated data model. After we acquire the stage 1 MCMC sample, we evaluate the conditional PMF $[n|\boldsymbol\beta,\psi]$ for all realizations of $\boldsymbol\beta$ and $\psi$ from the first stage in parallel.  This intermediate step requires numerical integration \citep[e.g.,][]{bonner2014mc} to approximate (\ref{eq:npartialintSCR}) and thus parallelization improves computational time substantially.

In stage 2, we use PPRB to update the model parameters using joint random draws from the first stage MCMC sample as proposals $\{\boldsymbol\beta^{(*)}, \psi^{(*)} \}$ and the Metropolis-Hastings ratio
\begin{equation}
  r=\frac{[n|\boldsymbol\beta^{(*)},\psi^{(*)}]}{[n|\boldsymbol\beta^{(k-1)},\psi^{(k-1)}]} \;. 
  \label{eq:SCR_mh}
\end{equation}
This second stage can be performed quickly using a look-up table for the pre-computed conditional PMFs for $n$ resulting from the first stage.  

We follow the procedure described in the previous section to obtain a MCMC sample for population abundance $N$ in a third computing stage.  Under the SCR model based on a conditional Poisson assumption for $n$, we sample $N_0^{(k)}$ as     
\begin{equation}
  N_0^{(k)} \sim \text{Pois}(\bar{\psi}^{(k)}(M-n)) \;,
  \label{eq:scrN0sample}
\end{equation}
and then let $N^{(k)}=n+N_0^{(k)}$ for $k=1,\ldots,K$ second-stage MCMC iterations.  This form of full-conditional updating is possible because, after we condition on $n$, we know that the additional undetected individuals ($N_0$) from our population are indistinguishable and independent with full-conditional membership probability    
\begin{equation}
  \bar{\psi}^{(k)}=\int \left(\frac{\psi^{(k)}\prod_{l=1}^L(1-p^{(k)}_l)^J}{\psi^{(k)}\prod_{l=1}^L(1-p^{(k)}_l)^J+1-\psi^{(k)}}\right) [\mathbf{s}]d\mathbf{s} \;,
  \label{eq:SCRpsibar}
\end{equation}
\noindent where $p^{(k)}_l=\text{logit}^{-1}(\beta_0^{(k)}+\beta_1^{(k)}||\mathbf{s}-\mathbf{x}_l||^2_2)$ for $k=1,\ldots,K$ second-stage MCMC iterations.  Furthermore, the sampling of $N_0$ can be performed in parallel \emph{post hoc} because all quantities in the full-conditional distribution (\ref{eq:scrN0sample}) have already been obtained in the second computing stage.

We can calculate the posterior power to detect for the SCR model using a similar approach as described in the previous section.  However, because we have collected CR data across an entire trap array, the probability of detecting a random individual in $J$ sampling occasions is
\begin{equation}
    \text{Pr}\left(\sum_{l=1}^L\tilde{y}_l>0\Big\vert\tilde{z}=1,\mathbf{Y}\right)=\int \mathbbm{1}_{\{\sum_{l=1}^L\tilde{y}_l>0\}} [\tilde{\mathbf{y}}|\tilde{z}=1,\mathbf{Y}]d\tilde{\mathbf{y}} \;,
  \label{eq:ptd_scr}
\end{equation}
\noindent which involves a multidimensional integral, but can still be approximated using composition sampling to obtain a posterior predictive MCMC sample for $\tilde{y}_l^{(k)}$ for $l=1,\ldots,L$ and $k=1,\ldots,K$.  Monte Carlo integration can then be used to calculate the posterior power to detect as $\text{Pr}\left(\sum_{l=1}^L\tilde{y}_l>0\big\vert\tilde{z}=1,\mathbf{Y}\right)=\sum_{k=1}^K \mathbbm{1}_{\{\sum_{l=1}^L \tilde{y}^{(k)}_l>0\}}/K$.  We also report $E(n/N|\mathbf{Y})$ for both models as an alternative way to infer the power to detect.      
 
\subsection{Application:  Snowshoe hare abundance}
\begin{figure}[htp]
  \centering
  \includegraphics[width=6in]{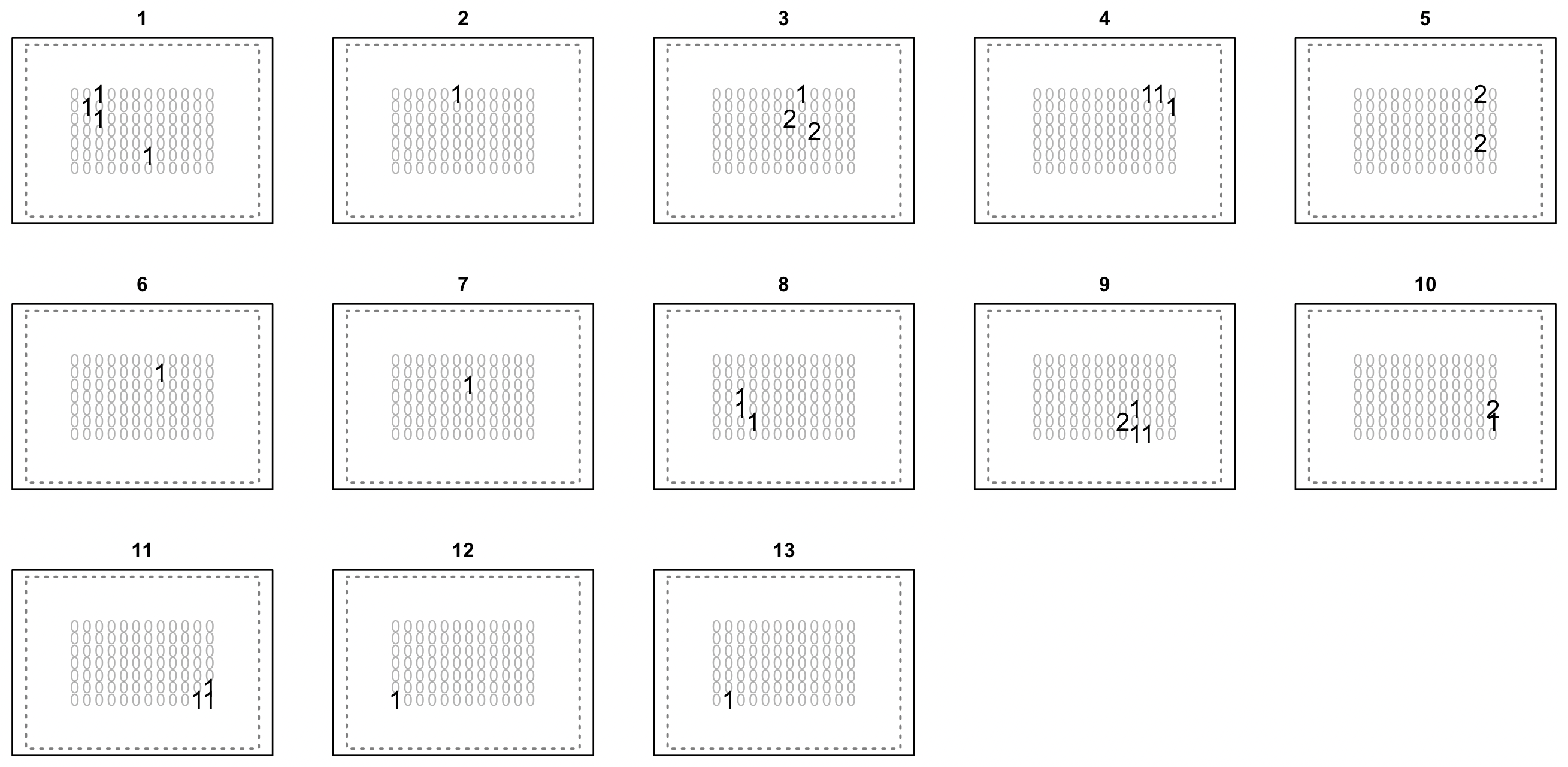}
  \caption{SCR data for $n=13$ snowshoe hare individuals over a $7\times 12$ array of $L=84$ traps spaced 50m apart.  Positions of numbers represent trap locations in array and values correspond to the number of detections for each individual at each trap (cases with $y_{i,l}>0$ shown in bold).  Support for activity centers $\mathbf{s}$ is shown as a dashed box; expanded 200 m in each direction from the extent of the trap array.}
  \label{fig:SCR_data}
\end{figure}

We demonstrate the PPRB approach to implementing the SCR model using a data set comprised of encounter histories of $n=13$ snowshoe hares (\emph{Lepus} \emph{americanus}) based on $J=5$ sampling occasions at an array of $L=84$ traps in central Colorado, USA \citep{ivan2014density} during winter 2007.  The snowshoe hare SCR data are shown in Figure~\ref{fig:SCR_data} on a regular grid of trap locations spaced 50m apart at which individuals were captured with live traps and marked using passive integrated transponder tags that identified individuals on recapture (data in Appendix C).     

We fit the SCR model with both the Poisson-binomial and Poisson conditional distributions for $n$ to the data shown in Figure~\ref{fig:SCR_data}.  For priors, we specified $\boldsymbol\beta\sim \text{N}(\mathbf{0},1000\cdot\mathbf{I})$, $\psi\sim\text{Beta}(1,1)$, and $\mathbf{s}_i\sim \text{Unif}({\cal A})$, where ${\cal A}$ is a rectangular region extending 200 m beyond the trap array in each direction (dashed region in Figure~\ref{fig:SCR_data}; 66.5ha area).  We assumed a total of $M=200$ in the superpopulation, which implies $M-n=187$ augmented individuals with all-zero encounter histories.  However, similar to our implementation of the heterogeneous model in the previous section, we do not actually augment the data set to fit this SCR model.  Instead, $M$ is involved in computing the components (\ref{eq:npartialintSCR}) of the second stage Metropolis-Hastings ratios (\ref{eq:SCR_mh}).  

We fit the models using $K=100000$ MCMC iterations, which required 117.75 minutes for stage one.  Stage two (including the intermediate parallel stage to evaluate the partially integrated full-conditional distributions for $n$) in the Poisson-binomial case required 5.24 minutes and 1.36 minutes total for the Poisson case.  For comparison, a single-stage algorithm (in JAGS) required $2.81$ minutes per effective sample, whereas the Poisson-binomial and Poisson recursive algorithms only required $0.021$ and $0.019$ minutes per effective sample respectively.  For this SCR model, the recursive algorithms were substantially faster than the single-stage algorithm relative to effective sample size, but similar among themselves (although the Poisson case was 4 times faster than the Poisson-binomial per raw iteration).  These computing speed characteristics are a function of the number of augmented individuals ($M-n$) and power to learn the unknown parameters given the available data.  

\begin{figure}[htp]
  \centering
  \includegraphics[width=5in]{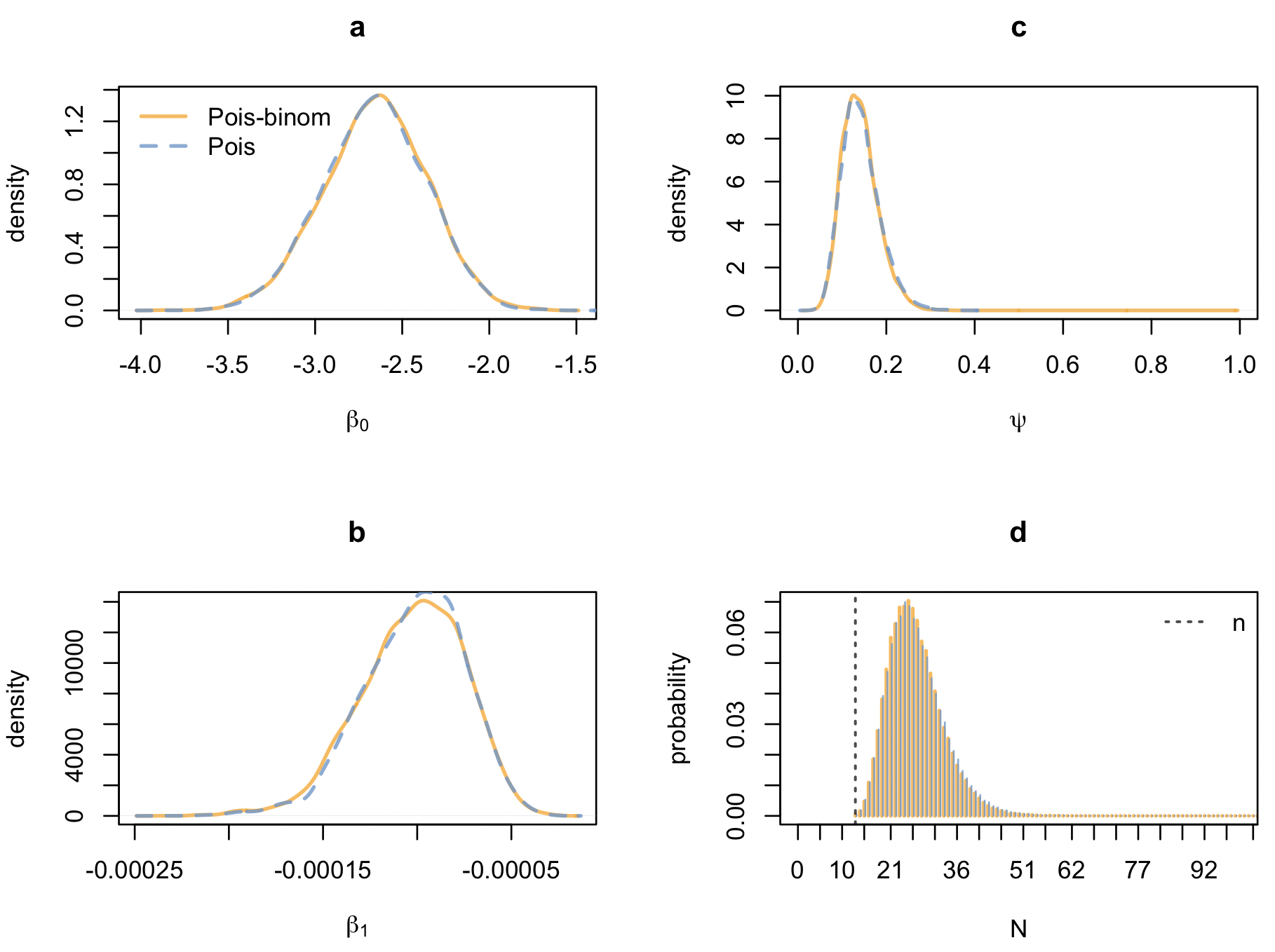}
  \caption{Marginal posterior distributions for a) $\beta_0$, b) $\beta_1$, c) $\psi$, and d) $N$.  Distributions shown are a result of the multistage Bayesian algorithm used to fit the SCR model with conditional Poisson-binomial (orange) and Poisson (blue) assumption for $n$.  Subfigure d shows marginal posterior probability mass functions for $N$.}
  \label{fig:SCR_results}
\end{figure}

\begin{figure}[htp]
  \centering
  \includegraphics[width=5in]{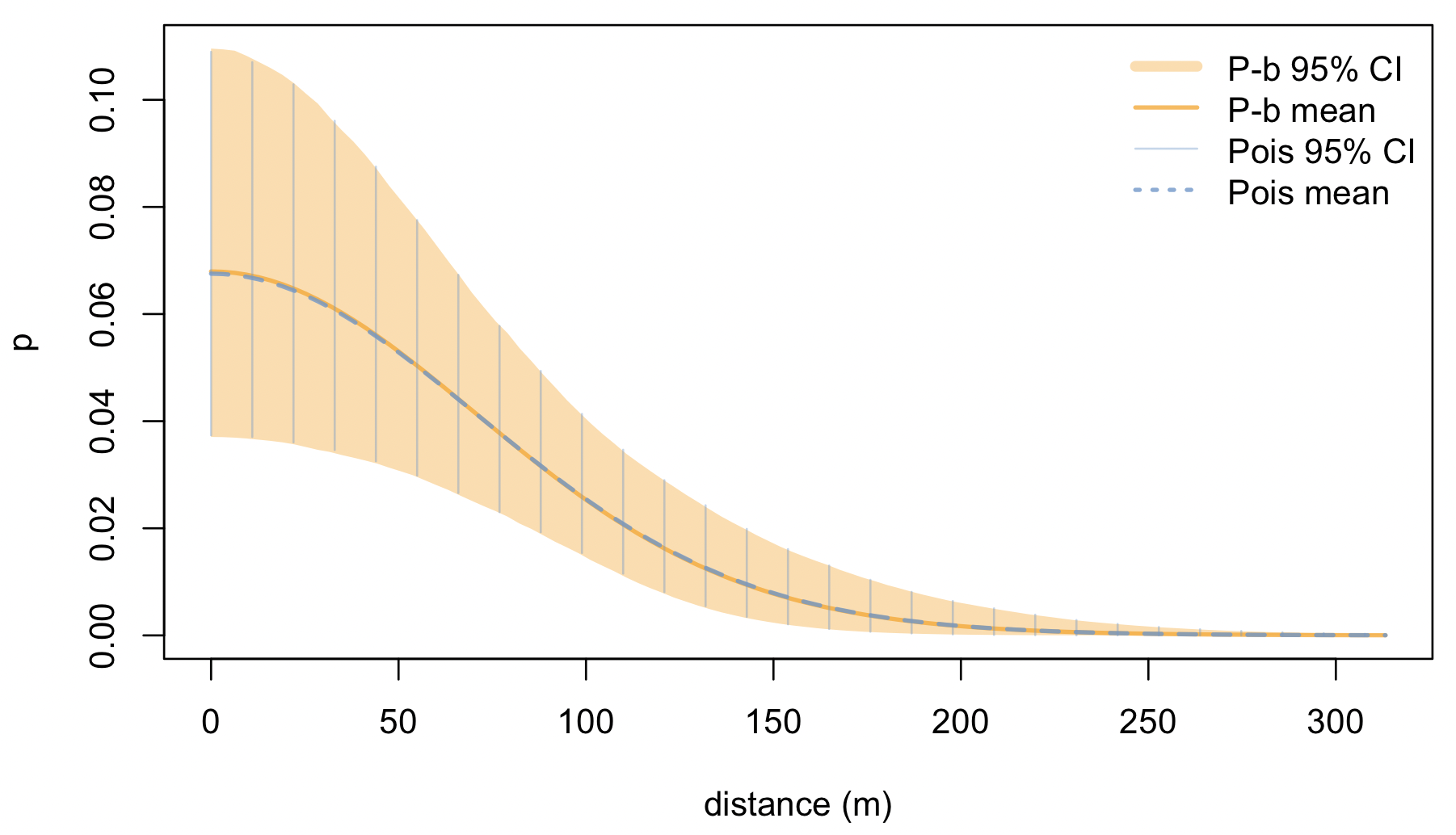}
  \caption{Pointwise 95\% credible intervals (shaded and hashed) and posterior means (solid and dashed) for the detection function $p$ based on fitting the SCR model with Poisson-binomial (P-b; orange) and Poisson (Pois; blue) conditional models for $n$.}
  \label{fig:SCR_p}
\end{figure}

The posterior results from fitting the SCR model with conditional Poisson-binomial and Poisson assumptions for $n$ to the snowshoe hare data are summarized in Figure~\ref{fig:SCR_results}. 
Our results indicate that the capture probabilities of snowshoe hares are small in general (i.e., $E(p|\mathbf{Y})\approx 0.07$ for a trap placed at the individual activity center using either model specification) and they decrease away from the activity centers as we expect given the space use mechanism associated with the SCR model (e.g., $E(p|\mathbf{Y})\approx 0.025$ for a trap 100m away from the individual activity center using either model specification).  In fact, Figure~\ref{fig:SCR_p} shows the estimated detection function associated with our SCR models over a range of distances spanning half the maximum distance in the study trap array. These results indicate the two models are very similar and both with very low probability of detecting an individual with a trap farther than 200 m from the individual's activity center (i.e., outside our study area).   

We obtained a posterior sample for abundance $N$ as described previously, by computing $\bar\psi^{(k)}$, sampling $N_0^{(k)}$ in parallel, and letting $N^{(k)}=n+N_0^{(k)}$.  Based on the Poisson-binomial model, we estimated posterior mean abundance as $E(N|\mathbf{Y})=26.6$ with a 95\% posterior credible interval of $(17,40)$.  The Poisson model resulted in $E(N|\mathbf{Y})=26.8$ with a 95\% posterior credible interval of $(17,42)$. Thus, despite the low detection probability, with $J=5$ sampling occasions, approximately half of the individuals in the population were observed ($n=13$) with only a few individuals possibly going undetected in our study area.  In fact, the posterior power to detect for both models was approximately $\text{Pr}\left(\sum_{l=1}^L\tilde{y}_l>0|\tilde{z}=1,\mathbf{Y}\right)=0.51$ based on $J=5$ sampling occasions for a random individual from our population in the study area (dashed region in Figure~\ref{fig:SCR_data}).  The alternative power to detect (i.e., $E(n/N|\mathbf{Y})$) was $0.51$ for both forms of the model.      

\section{Discussion}
A variety of approaches to formulating and implementing CR models have been developed since early work in this area in the mid-20th century \citep{schofield201650}.  Many modern implementations of CR models are Bayesian, and a substantial portion of those rely on some form of data augmentation strategy \citep{durban2005mark, royle2012parameter}.  The data augmentation perspective is intuitive and facilitates model generalizations that can account for real-world complexities such as the effect of animal space use patterns on the detection function associated with an array of traps.  However, as these data sets grow in size and variety, conventional algorithms to fit complicated CR models to augmented data sets may not be computationally efficient \citep{yackulic2020need}.  

Following \cite{king2016capture}, we showed how to reformulate a large class of CR models in a way that is based on the intuitive PX-DA framework.  We then showed how to fit them using multistage computing strategies.  The natural partitioning of observed versus augmented data illuminates an explicit conditioning on the number of observed individuals $n$, which, in turn, has its own conditional model that depends on parameters and is implied by the PX-DA scheme.  Similar model specifications have been derived from alternative perspectives \citep[e.g.,][]{borchers2008spatially,king2016capture} but have not been leveraged to facilitate multistage Bayesian computing strategies.  

Conditional perspectives are not new in the analysis of CR data \citep[e.g.,][]{sanathanan1972estimating,huggins1989statistical,huggins1991some,worthington2015analysing,king2016capture}.  However, writing CR models as a product of a conditional distribution for the detections and a distribution for sample size allows for new model formulations.  For example, in cases where individuals may be clustered in the population due to family groups or social structure, it may be advantageous to specify an overdispersed count model such as a negative binomial or quasi-Poisson \citep[e.g.,][]{ver2007quasi}.  Conversely, if individuals are more regularly distributed in the population due to mechanisms such as territoriality, an underdispersed count model such as a Conway-Maxwell Poisson could be specified \citep[e.g.,][]{shmueli2005useful}.  These model formulations involve additional parameters and may benefit from additional data, but our conditional strategy for representing the model provides a way to accommodate these extra sources of dependence in future studies. 

Using a recursive implementation, we showed how to fit CR models in two stages.  The first computing stage fits a zero-truncated CR model to the observed data (of dimension $n$ only).  We then resample the first-stage output based on a secondary algorithm that assimilates the sample size information $n$.  For MCMC specifically, calculating the necessary ratios in the second stage can be a computing bottleneck because it requires numerical integration.  However, we can accelerate the second computing stage substantially by evaluating the components in parallel between stages.  Thus, the first-stage only involves $n$ observations, the intermediate parallel computation scales with the number of available cores, and the second stage only requires a look-up table to compute the necessary ratios.  Inference for abundance $N$ can be obtained after model fitting, where the undetected number of individuals $N_0$ is sampled from its full-conditional distribution.  This final step can also be parallelized to reduce computation time.    

Our application of recursive computing techniques to fit CR models aligns well with other recent developments, including multiple imputation and the explicit consideration of ancillarity in these types of models \citep{worthington2015analysing,schofield201650}.  A promising area of future research should seek to formally connect the ancillarity concepts with recursive computing strategies.  Our approach also aligns well with other calls for Rao-Blackwellization (i.e., marginalization) in the Bayesian implementation of CR models \citep[e.g.,][]{yackulic2020need}.  One additional benefit of the multistage computing approach to implementing these models is that the first computing stage can be performed using automatic Bayesian software that may incorporate adaptive tuning techniques and alleviate the need for supervised MCMC algorithms altogether. 

For completeness, we note that a variety of other approaches exist for implementing CR models, including Dirichlet process approaches \citep{manrique2016bayesian,diana2020hierarchical}, numerical integration \citep{coull1999use,borchers2008spatially}, and transdimensional methods like reversible-jump MCMC \citep{king2008bayesian,mclaughlin2019}.  In fact, \cite{king2016capture} noted that working with the integrated likelihood based on the PX-DA model formulation may be sufficient for fitting certain classes of CR models.  Thus, recursive computing strategies may not always be necessary.  However, our focus is to illuminate alternative ways to specify CR models that may not have been apparent otherwise.  The resulting models themselves could be implemented using a variety of computing strategies depending on the goals and constraints of the study. 

\section*{Acknowledgments} 
This research was funded by NSF DEB 1927177.  The authors thank Larissa Bailey for providing data, Matt Schofield and Richard Barker for helpful conversations, and anonymous reviewers and associate editor for thoughtful suggestions.  


\bibliographystyle{biom} 
\bibliography{cr_refs}


\newpage
\pagestyle{empty} 
\begin{center}
 \Large Appendices for ``Multistage Hierarchical Capture-Recapture Models''
\end{center}
\subsection*{Appendix A}
To fit the homogeneous CR model using a single-stage MCMC algorithm, we consider the joint posterior distribution for $p$ and $\psi$.  Under the PX-DA framework, this posterior distribution can be written as
\begin{align}
  [p,\psi|\mathbf{y}_{1:n},\mathbf{y}_{(n+1):M},n]&\propto [\mathbf{y}_{(n+1):M}|p,\psi,\mathbf{y}_{1:n},n][p,\psi|\mathbf{y}_{1:n},n]\;, \\
  &\propto [\mathbf{y}_{(n+1):M}|p,\psi,\mathbf{y}_{1:n},n][\mathbf{y}_{1:n}|p,n][n|p,\psi][p][\psi] \;, \\ 
  &\propto [\mathbf{y}_{1:n}|p,n][n|p,\psi][p][\psi] \;, 
  \label{eq:condpost_app}
\end{align}
\noindent where the full-conditional distribution of $\mathbf{y}_{(n+1):M}$ is proportional to one when conditioned on $n$ (and hence drops out of the right hand side) and the conditional distribution of $\mathbf{y}_{1:n}$ is proportional to the product of zero-truncated binomials
\begin{equation}
  [\mathbf{y}_{1:n}|p,n] \propto \frac{\prod_{i=1}^n [y_i|p]}{(1-(1-p)^J)^n} \;.
\end{equation}
For a given joint proposal distribution $[p,\psi]^*$, the associated Metropolis-Hastings ratio to update $p$ and $\psi$ jointly is
\begin{equation}
  r=\frac{[\mathbf{y}_{1:n}|p^{(*)},n][n|p^{(*)},\psi^{(*)}][p^{(*)}][\psi^{(*)}][p^{(k-1)},\psi^{(k-1)}]^*}{[\mathbf{y}_{1:n}|p^{(k-1)},n][n|p^{(k-1)},\psi^{(k-1)}][p^{(k-1)}][\psi^{(k-1)}][p^{(*)},\psi^{(*)}]^*} \;.
\end{equation}

To implement the model using PPRB following \cite{hooten2019making}, we obtain an initial MCMC sample for $p$ and $\psi$ by fitting the CR model to the observed data while conditioning on fixed and known $n$.  The posterior distribution for the first stage is proportional to
\begin{equation}
  [\mathbf{y}_{1:n}|p,n][p][\psi] \;,
\end{equation}
\noindent with respect to $p$ and $\psi$, and the associated first-stage Metropolis-Hastings ratio is 
\begin{equation}
  r=\frac{[\mathbf{y}_{1:n}|p^{(*)},n][p^{(*)}][\psi^{(*)}][p^{(k-1)},\psi^{(k-1)}]^*}{[\mathbf{y}_{1:n}|p^{(k-1)},n][p^{(k-1)}][\psi^{(k-1)}][p^{(*)},\psi^{(*)}]^*} \;.
\end{equation}
At this first stage, we use a temporary proposal distribution $[p,\psi]^*$ that is convenient.  

For the second stage of the PPRB implementation, we assume that the proposal distribution is
\begin{equation}
  [p,\psi]^*\propto [\mathbf{y}_{1:n}|p,n][p][\psi] \;,
\end{equation}
\noindent which is equivalent to the first-stage posterior, and randomly sample (with replacement) joint first-stage MCMC realizations to use as proposals in the second stage Metropolis-Hastings updates.  The resulting second-stage Metropolis-Hastings ratio becomes   
\begin{align}
  r&=\frac{[\mathbf{y}_{1:n}|p^{(*)},n][n|p^{(*)},\psi^{(*)}][p^{(*)}][\psi^{(*)}][p^{(k-1)},\psi^{(k-1)}]^*}{[\mathbf{y}_{1:n}|p^{(k-1)},n][n|p^{(k-1)},\psi^{(k-1)}][p^{(k-1)}][\psi^{(k-1)}][p^{(*)},\psi^{(*)}]^*} \;, \\
  &=\frac{[n|p^{(*)},\psi^{(*)}]}{[n|p^{(k-1)},\psi^{(k-1)}]} \;,
\end{align}
\noindent because the proposal cancels with the data model and priors.  Thus, the second-stage Metropolis-Hastings ratio is merely a quotient involving the conditional model for $n$ and can be evaluated easily using the first-stage MCMC sample.     

\subsection*{Appendix B}
To demonstrate the recursive implementation of the hierarchical CR model, we fit the model to simulated data using the single-stage and two-stage approaches.  To simulate CR data for this example, we set $M=100$ individuals in the superpopulation, membership probability $\psi=0.4$, and detection probability $p=0.25$.  Then we used the hierarchical model in Section 2 as a generative process to simulate data based on $J=3$ occasions which resulted in $N=39$ individuals in our population, with $n=19$ individuals observed by our measurement process.  The simulated observed data $y_i$, for $i=1,\ldots,n$, can be summarized by the values $\mathbf{y}=(1,1,1,1, 1,1,1,2,2,1,1,1,1,2,2,1,1,2,1)'$.   

In the PX-DA implementation, we assumed $M-n=81$ augmented individuals with all-zero capture histories.  Thus, we assumed the same $M$ as in our data simulation and this allows us to infer the true $\psi$.  We note that the estimation of $N$ is not constrained by $M$ empirically in this example, therefore we could use larger values of $M$ without influencing the inference for parameter $p$.   

\begin{figure}[htp]
  \centering
  \includegraphics[width=4in]{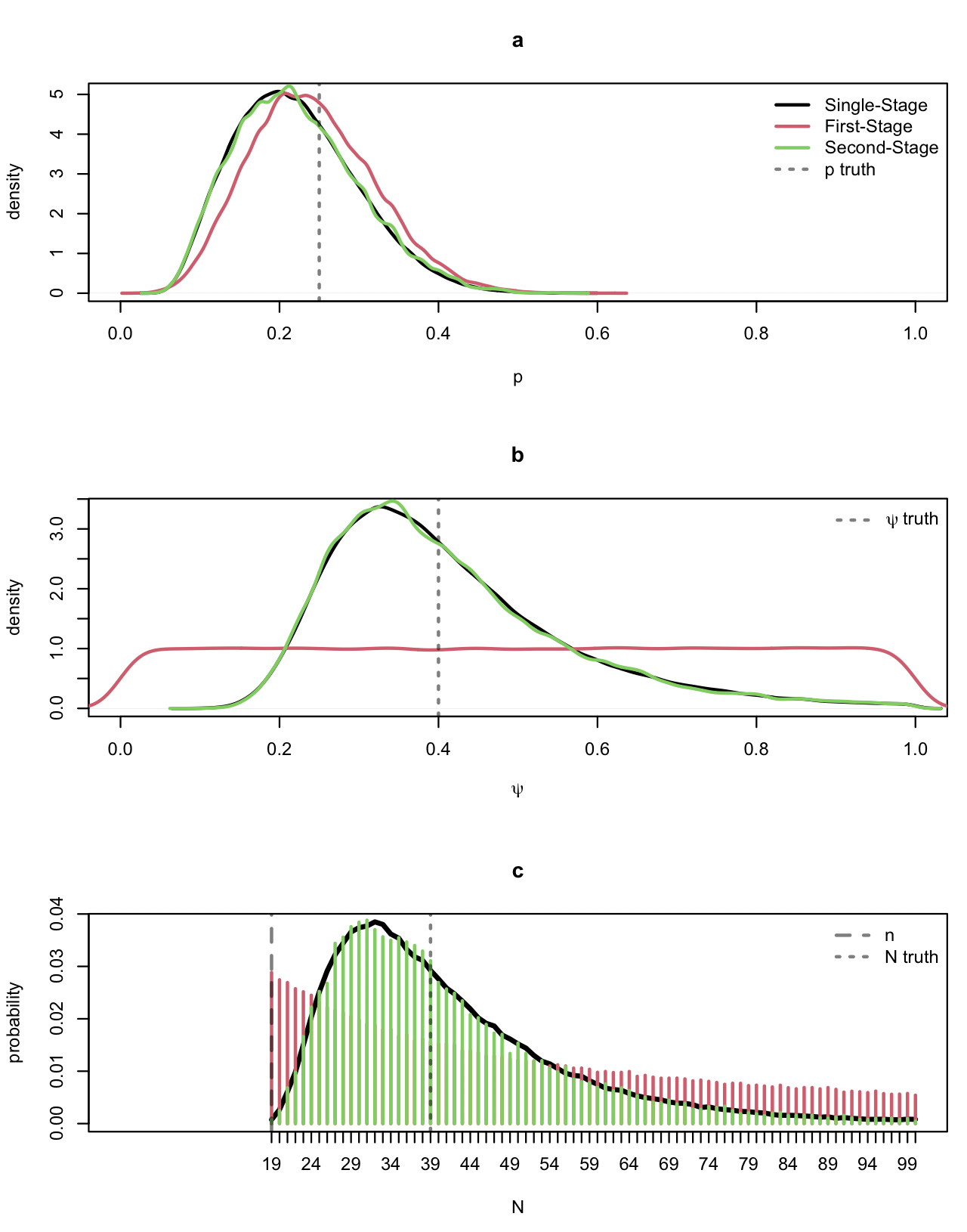}
  \caption{Marginal posterior distributions for a) $p$, b) $\psi$, c) $N$.  Distributions shown are a result of the single-stage MCMC algorithm (black), first-stage of the two-stage MCMC algorithm (red), and second-stage of the two-stage MCMC algorithm (green).  Subfigure c shows marginal posterior probability mass functions; black line shown for single-stage for comparison.}
  \label{fig:M0_results}
\end{figure}

We fit the hierarchical CR from Section 2 to our simulated data using two approaches:  1) a standard single-stage MCMC algorithm for the hierarchical model and 2) a two-stage algorithm based on the recursive formulation of the same model.  In both cases, we used $K=200000$ MCMC iterations.  The results of our analyses are summarized in Figure~\ref{fig:M0_results}.  
The posterior comparison shown in Figure~\ref{fig:M0_results} indicates that the two-stage PPRB approach yields the same inference as the conventional single-stage MCMC algorithm.  Both approaches fit exactly the same model, but the recursive framework suggests that other specifications for the conditional model for $n$ (e.g., Poisson) are straightforward to implement.  Furthermore, in more complicated models, we can benefit from parallel evaluation of the PMF for $n$ which can improve stability and facilitate computation for large data sets.        

\pagebreak
\subsection*{Appendix C}
The snowshoe hare SCR data used to fit the model in Section 4 are presented below.  The first column is the trap ID (i.e., $[1,]$ indicates trap 1), the second two columns correspond to the $L=84$ trap locations $\mathbf{X}$ in meters, and the remaining $13$ columns contain values that correspond to the sum of detections associated with each individual at each trap.
\tiny
\begin{verbatim}
 [1,]    0    0 0 0 0 0 0 0 0 0 0 0 0 0 0
 [2,]   50    0 0 0 0 0 0 0 0 0 0 0 0 0 0
 [3,]  100    0 1 0 0 0 0 0 0 0 0 0 0 0 0
 [4,]  150    0 0 0 0 0 0 0 0 0 0 0 0 0 0
 [5,]  200    0 0 0 0 0 0 0 0 0 0 0 0 0 0
 [6,]  250    0 0 1 0 0 0 0 0 0 0 0 0 0 0
 [7,]  300    0 0 0 0 0 0 0 0 0 0 0 0 0 0
 [8,]  350    0 0 0 1 0 0 0 0 0 0 0 0 0 0
 [9,]  400    0 0 0 0 0 0 0 0 0 0 0 0 0 0
[10,]  450    0 0 0 0 1 0 0 0 0 0 0 0 0 0
[11,]  500    0 0 0 0 1 2 0 0 0 0 0 0 0 0
[12,]  550    0 0 0 0 0 0 0 0 0 0 0 0 0 0
[13,]    0  -50 0 0 0 0 0 0 0 0 0 0 0 0 0
[14,]   50  -50 1 0 0 0 0 0 0 0 0 0 0 0 0
[15,]  100  -50 0 0 0 0 0 0 0 0 0 0 0 0 0
[16,]  150  -50 0 0 0 0 0 0 0 0 0 0 0 0 0
[17,]  200  -50 0 0 0 0 0 0 0 0 0 0 0 0 0
[18,]  250  -50 0 0 0 0 0 0 0 0 0 0 0 0 0
[19,]  300  -50 0 0 0 0 0 0 0 0 0 0 0 0 0
[20,]  350  -50 0 0 0 0 0 1 0 0 0 0 0 0 0
[21,]  400  -50 0 0 0 0 0 0 0 0 0 0 0 0 0
[22,]  450  -50 0 0 0 0 0 0 0 0 0 0 0 0 0
[23,]  500  -50 0 0 0 0 0 0 0 0 0 0 0 0 0
[24,]  550  -50 0 0 0 1 0 0 0 0 0 0 0 0 0
[25,]    0 -100 0 0 0 0 0 0 0 0 0 0 0 0 0
[26,]   50 -100 0 0 0 0 0 0 0 0 0 0 0 0 0
[27,]  100 -100 1 0 0 0 0 0 0 0 0 0 0 0 0
[28,]  150 -100 0 0 0 0 0 0 0 0 0 0 0 0 0
[29,]  200 -100 0 0 0 0 0 0 0 0 0 0 0 0 0
[30,]  250 -100 0 0 0 0 0 0 0 0 0 0 0 0 0
[31,]  300 -100 0 0 2 0 0 0 1 0 0 0 0 0 0
[32,]  350 -100 0 0 0 0 0 0 0 0 0 0 0 0 0
[33,]  400 -100 0 0 0 0 0 0 0 0 0 0 0 0 0
[34,]  450 -100 0 0 0 0 0 0 0 0 0 0 0 0 0
[35,]  500 -100 0 0 0 0 0 0 0 0 0 0 0 0 0
[36,]  550 -100 0 0 0 0 0 0 0 0 0 0 0 0 0
[37,]    0 -150 0 0 0 0 0 0 0 0 0 0 0 0 0
[38,]   50 -150 0 0 0 0 0 0 0 0 0 0 0 0 0
[39,]  100 -150 0 0 0 0 0 0 0 1 0 0 0 0 0
[40,]  150 -150 0 0 0 0 0 0 0 0 0 0 0 0 0
[41,]  200 -150 0 0 0 0 0 0 0 0 0 0 0 0 0
[42,]  250 -150 0 0 0 0 0 0 0 0 0 0 0 0 0
[43,]  300 -150 0 0 0 0 0 0 0 0 0 0 0 0 0
[44,]  350 -150 0 0 0 0 0 0 0 0 0 0 0 0 0
[45,]  400 -150 0 0 2 0 0 0 0 0 0 0 0 0 0
[46,]  450 -150 0 0 0 0 0 0 0 0 0 0 0 0 0
[47,]  500 -150 0 0 0 0 0 0 0 0 0 0 0 0 0
[48,]  550 -150 0 0 0 0 0 0 0 0 0 0 0 0 0
[49,]    0 -200 0 0 0 0 0 0 0 0 0 0 0 0 0
[50,]   50 -200 0 0 0 0 0 0 0 0 0 0 0 0 0
[51,]  100 -200 0 0 0 0 0 0 0 1 0 0 0 0 0
[52,]  150 -200 0 0 0 0 0 0 0 0 0 0 0 0 0
[53,]  200 -200 0 0 0 0 0 0 0 0 0 0 0 0 0
[54,]  250 -200 0 0 0 0 0 0 0 0 0 0 0 0 0
[55,]  300 -200 0 0 0 0 0 0 0 0 0 0 0 0 0
[56,]  350 -200 0 0 0 0 0 0 0 0 0 0 0 0 0
[57,]  400 -200 0 0 0 0 0 0 0 0 1 0 0 0 0
[58,]  450 -200 0 0 0 0 0 0 0 0 0 0 0 0 0
[59,]  500 -200 0 0 0 0 2 0 0 0 0 0 0 0 0
[60,]  550 -200 0 0 0 0 0 0 0 0 0 2 0 0 0
[61,]    0 -250 0 0 0 0 0 0 0 0 0 0 0 0 0
[62,]   50 -250 0 0 0 0 0 0 0 0 0 0 0 0 0
[63,]  100 -250 0 0 0 0 0 0 0 0 0 0 0 0 0
[64,]  150 -250 0 0 0 0 0 0 0 1 0 0 0 0 0
[65,]  200 -250 0 0 0 0 0 0 0 0 0 0 0 0 0
[66,]  250 -250 0 0 0 0 0 0 0 0 0 0 0 0 0
[67,]  300 -250 1 0 0 0 0 0 0 0 0 0 0 0 0
[68,]  350 -250 0 0 0 0 0 0 0 0 2 0 0 0 0
[69,]  400 -250 0 0 0 0 0 0 0 0 0 0 0 0 0
[70,]  450 -250 0 0 0 0 0 0 0 0 0 0 0 0 0
[71,]  500 -250 0 0 0 0 0 0 0 0 0 0 0 0 0
[72,]  550 -250 0 0 0 0 0 0 0 0 0 1 1 0 0
[73,]    0 -300 0 0 0 0 0 0 0 0 0 0 0 1 0
[74,]   50 -300 0 0 0 0 0 0 0 0 0 0 0 0 1
[75,]  100 -300 0 0 0 0 0 0 0 0 0 0 0 0 0
[76,]  150 -300 0 0 0 0 0 0 0 0 0 0 0 0 0
[77,]  200 -300 0 0 0 0 0 0 0 0 0 0 0 0 0
[78,]  250 -300 0 0 0 0 0 0 0 0 0 0 0 0 0
[79,]  300 -300 0 0 0 0 0 0 0 0 0 0 0 0 0
[80,]  350 -300 0 0 0 0 0 0 0 0 0 0 0 0 0
[81,]  400 -300 0 0 0 0 0 0 0 0 1 0 0 0 0
[82,]  450 -300 0 0 0 0 0 0 0 0 1 0 0 0 0
[83,]  500 -300 0 0 0 0 0 0 0 0 0 0 1 0 0
[84,]  550 -300 0 0 0 0 0 0 0 0 0 0 1 0 0
\end{verbatim}

\end{document}